# The Cryogenic System for the Panda-X Dark Matter Search Experiment

Zhao Li[1], Karl Ludwig Giboni[1], Haowei Gong[1], Xiangdong Ji[1,2], and Andy Tan[1]

[1]*INPAC, Department of Physics, and Shanghai Key Laboratory for Particle Physics and Cosmology*
*Shanghai Jiao Tong University, Shanghai 200240, P.R. China*
[2]*Department of Physics, University of Maryland, College Park, MD 20742, USA*
E-mail: xdji@sjtu.edu.cn

ABSTRACT: Panda-X is a liquid xenon dual-phase detector for the Dark Matter Search. The first modestly-sized module will soon be installed in the China JinPing Deep Underground Laboratory in Sichuan province, P.R. China. The cryogenics system is designed to handle much larger detectors, even the final version in the ton scale. Special attention has been paid to the reliability, serviceability, and adaptability to the requirements of a growing experiment. The system is cooled by a single Iwatani PC150 Pulse Tube Refrigerator. After subtracting all thermal losses, the remaining cooling power is still 82W. The fill speed was 9 SLPM, but could be boosted by LN2 assisted cooling to 40 SLPM. For the continuous recirculation and purification through a hot getter, a heat exchanger was employed to reduce the required cooling power. The recirculation speed is limited to 35 SLPM by the gas pump. At this speed, recirculation only adds 18.5 W to the heat load of the system, corresponding to a 95.2 % efficiency of the heat exchanger.

**Contents**

1. **Introduction**

Liquid xenon presently enjoys great popularity for its use in large-size Dark Matter experiments. After the successes of XENON10[1] and XENON100[2] in the Gran Sasso National Laboratory in Italy, several other projects have aimed at developing even larger and more sensitive detectors in the near future. Among these is the Panda-X [ref. 3] experiment detector, which will be installed in the China JinPing Deep Underground Laboratory (CJPL) in Sichuan province with an overburden about 2500 m of rock (nearly 7000 meter water equivalent).

Although Panda-X will initially have only a modest 25 kg active target, it can house a very large liquid xenon mass in preparation for a larger future detector. The present detector is the first step in a multi-step development plan. The entire infrastructure in the underground lab will be dimensioned for the final ton scale. This means that the detector vessel, the external shield, and the cryogenic system must be capable to support a detector with about 1.5 ton of liquid xenon, compared to the presently used 400 kg.

The detector itself will be encapsulated in a double-walled cryostat for thermal insulation. The inner vessel is 0.75 m in diameter and 1.25 m high. Initially the inner vessels will be over dimensioned and appropriate fillers will be used to reduce the amount of costly xenon used during the experiments. In between the inner and outer vessels seven layers of aluminized Mylar foil in an insulation vacuum reduce the influx of heat into the cryogenic detector and thus reduce the required cooling power.

2. **The 'Cooling Bus' Structure**

For any large xenon Dark Matter detector, the radioactive background must be reduced to a minimum. Many auxiliary assemblies like valves, pumps, sensors, and the PTR are connected to the inner vessel of the detector. However, the radioactivity of these devices cannot be easily controlled or reduced to acceptable levels. Thus, they must be removed to the outside of the passive lead/polyethylene shield. Nearly all of the cryogenic system must act remotely on the detector from a distance of 1 m or more., because of a large cold finger, as used for Ge-detectors, would be impractical. Thus the principle of heat pipes was chosen similar to the XENON100 experiment[4].

In a heat pipe, an appropriate liquid is a boiled off at one end and recondensed at the other, with the gas streaming one way and the liquid running back. The heat transfer in this case is not by conduction, but by the streaming gas and liquid molecules.

The ideal liquid in our case is, of course, xenon. The envelope of the heat pipe is emulated by two concentric tubes, connected to the gas space above the liquid in the inner vessel at one end, and a cryo-cooler at the other. The effective thermal conductivity of such a system is much better than that in a solid copper rod.

A 2" diameter pipe with a vacuum jacket for heat insulation was chosen to transport the gas. The liquid can run back in a separate concentric 3/8" pipe. The liquid in this pipe is driven by gravity, and the pipe assembly is therefore mounted at a 5° angle. They are enclosed in an 8" vacuum pipe with 7 layers of aluminized Mylar foil to reduce radiative heat entering from the outside walls. The diameters of the pipes chosen were very large, so that they can also be used as vacuum connections to the inner vessel and the outer cryostat. For all the cryogenic and vacuum devices connected to the detector, only a single 8" diameter perforation of the shield is needed. The tube can be 'dog-legged', i.e. with an angle within the shielding material, but it might be sufficient that γ-rays from the outside cannot reach the detector in a straight line of sight.

While the principle of the connecting tubes sounds simple, the construction of 3 layers of concentric tubing with several connections to devices is rather complex. We therefore developed a design with independent modules connecting to a common system of concentric pipes, similar to that of a bus-structure in a computer system. It was therefore named the 'Cooling Bus'. Each module performs a separate function, like cooling, recirculation, etc.

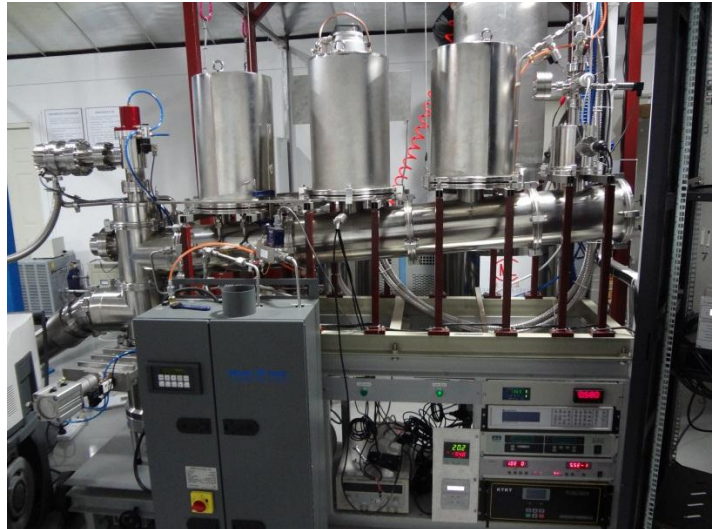

**Fig.1** The Cooling Bus. In front is the getter for the recirculation - purification.

All modules on the Cooling Bus have the same length and angle, and are equipped with the same flanges at the ends. The modules can now be chosen as required, and can be connected in any sequence or number. Figure 1 shows a photo of the Cooling Bus as used during the present test, and Figure 2 shows a schematic representation..

Just outside the shield the concentric tubes are connected to the Cooling Bus by a vertical section, which allows rotation around the center of the tubes. Thus, the Cooling Bus can be installed in nearly any direction. The vertical section is extended to the bottom via a 1200 l/sec turbo pump behind a 10" gate valve. The pump is for the insulation and cryostat vacuum. It is backed by a rotary vane pump. To the top, a 300 liters/sec turbo pump is mounted for the inner vessel. It can be separated before filling the detector with a 3" angle valve and is backed by a dry scroll pump. The valves are all actuated, so that they can be operated remotely.

## 3. PTR Module

The heart of the Panda-X cryogenic system is the Pulse Tube Refrigerator (PTR[5]). This PTR was originally designed[6] and optimized at KEK (The High Energy Accelerator Research Organization in Japan). Considering the tight space and the limited electric power in the JinPing lab we chose an air-cooled helium compressor[7]. This unit does not require a high capacity water chiller, which can be quite cumbersome in the limited space of an underground lab.

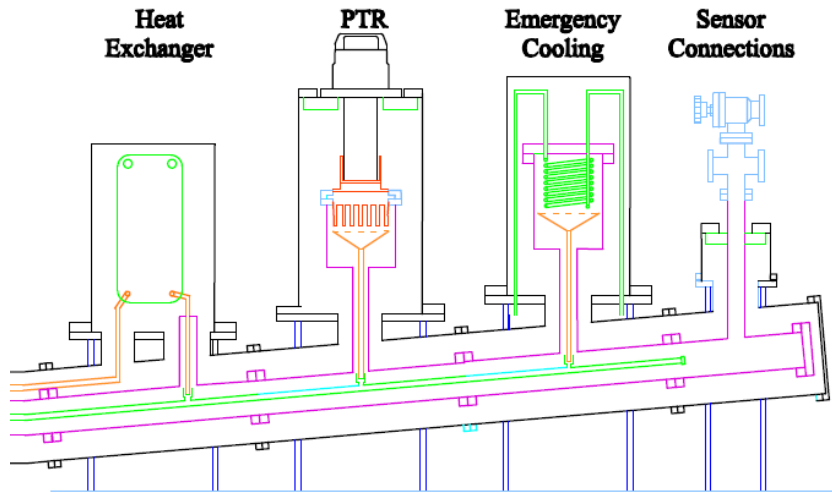

Fig 2 Schematic view of the Cooling Bus.

The cold head is equipped with a cup-shaped electrical heater and is mounted on an OFHC copper cold finger. Fins machined into the cold finger increase the surface to liquefy the xenon gas. The cold finger is sealed with an indium wire to a small cylindrical vessel which is connected to the 2" xenon pipe of the cooling bus. Thus, the PTR itself does not come in contact with the xenon; neither do the heating elements, the temperature sensors, or the connecting wires. The liquefied drops of liquefied xenon from the cold finger are collected into a funnel which guides them into the central liquid tube. This scheme of coupling the PTR indirectly to the xenon, i.e. via the cold finger, was first implemented in the XENON10 experiment[8]. It is thus possible to service or replace the PTR without opening the inner xenon vessel.

The temperature sensors are read by a temperature controller[9]. However, the maximum of heater of the controller is not sufficient to counteract the cooling power at all times. Therefore, the secondary control loop providing 0 - 10 V at low power controls a 300 W DC power supply.

The cooling power of the PTR in the present configuration was measured and optimized. For this purpose the inner system was evacuated, i.e. the system only presents a negligible heat load to the cryo cooler Figure 3 shows the cooling power (required heater power) versus cold head temperature. At -100°C the cooling power is 180 W, somewhat lower than the normal power of 200 W measured with a 7.5 kW Iwatani compressor. However, there are two effects which reduce the cooling power in our system. First, the compressor has a maximum consumption of 9 kW compared with the 7.5 kW for the Iwarani one, but it is aircooled, and 2 - 3 kW of it's power is used for cooling the compressor. Second, the Oxford Instruments is originally specified for 208 V 3-phase at 60Hz, whereas we operate the compressor at 190 V 3-phase at 50 Hz.

During most of the tests somewhat arbirarily an operating temperature of -94.5°C was chosen. This gives a good working pressure of about 1.2 barG, and also increases the distance from Triple Point, and reduces the probability of accidentally freezing the xenon. After filling the detector with liquid xenon at this temperature, the remaining cooling power was found to be 82 W, instead of 183 W measured with vacuum. This means the total heat entering the system is 101 W. The detector was filled with 100 kg of xenon, but we note that the required cooling power does not depend on the xenon mass, like in any system coupled with a heat pipe. For practical reasons, the bottom of the detector vessel did not have yet the Mylar foil, meaning that the surface (80 cm diameter) receives radiative heat from the outer cryostat vessel. For the next run, this remaining heat insulation will be in place.

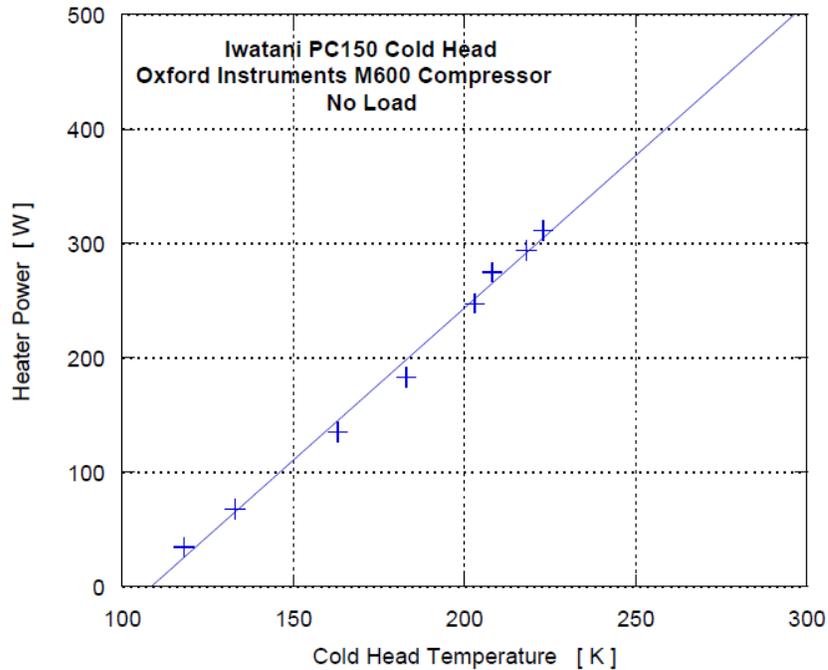

Fig. 3 Available cooling power (heater power) versus temperature for the PTR-compressor combination. The detector was evacuated and its heat load is negligible.

## 4. Continuous Recirculation - Purification

Constantly recirculating the liquid xenon through a high temperature getter gradually improves the purity and removes electro-negative substances out gassing from all surfaces of the detector. However, this does not mean the system has not to be thoroughly cleaned beforehand. Any remaining impurity concentrations determine the attenuation length for scintillation light and for drifting charges. Especially important are of course $H_2O$ and $O_2$. Naturally, the liquid has to be evaporated before entering the hot getter and recondensed afterward.

A large amount of thermal energy is stored in these phase transitions. For a flow of 1 SLPM (standard liter per minute) the power required to vary the temperature from -100°C to room temperature is about 1 W, but for the phase change about 10 W are needed. For a target flow rate of 100 SLPM one could use a heater of 1 kW, but the PTR would never be able to provide the cooling power to liquefy the purified xenon again. As proposed in a recent publication[10], we use a parallel-plate heat exchanger[11] to extract the heat from liquefying the gas and use it for evaporating the liquid going into the gas system. Fig. 3 shows a schematic of the recirculation loop. A double diaphragm pump[12] keeps the xenon streaming through the getter[13]. A flow controller is finally used to secure a constant flow. The pump had only a maximum delivery of 30 SLPM. It will be replaced with a larger model in the near future.

During a Dark Matter Search run the detector has typically to operate for an entire year with as few as possible interruptions. The need for redundancy valid for the cryo-cooler is also important for the recirculation system. Therefore, Panda-X will be equipped with two independent recirculation loops, each operating at 50 SLPM. For servicing one or the other can be decoupled from the system. Purifying with a single getter for a limited time will not affect the final purity, since the time driver in the process is the out gassing and not the ability of the getter to remove the impurities.

Testing the recirculation is very time consuming since for every operating point a new equilibrium has to be established in the detector. The approach to the equilibrium state has several time constants. After about 2 hrs the system seems to be stable, but it still changes significantly, but much slower during the next 12 hrs. All these tests were made at the same temperature setting of the PTR, -94.5°C, and the pressure was 1 BarG.

The maximum flow rate observed was 35 SLPM, limited by the pump delivery, not the cooling power of the PTR. We note that without heat exchanger about 350 W of cooling power would be needed at this rate. As discussed in ref. 10, the under pressure of the pump lets the xenon evaporate and cool itself. The slight difference in temperature is sufficient for the heat exchanger to function efficiently. Otherwise the phase changes would be at zero temperature difference, i.e. no heat transfer would occur.

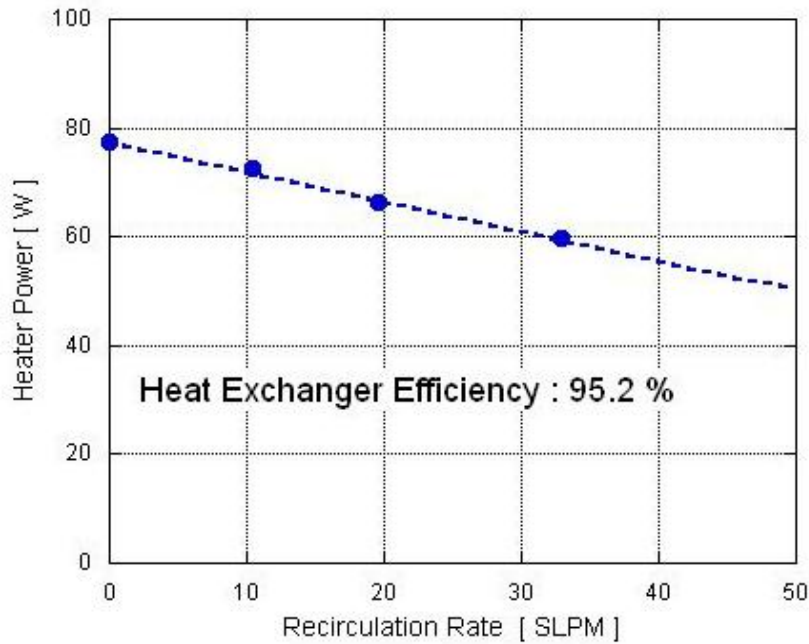

Fig. 4: Heater Power vs. Recirculation Flow Rate. The slope of the fitted line determines the inefficiency of the heat exchanger

Fig. 4 shows the flow rate versus the excess cooling power of the PTR, i.e. the power provided to heater above the cold finger. The measured points are lying on a straight line. The slope is interpreted as the efficiency of the heat exchanging process. It is 95.2 % is in very good agreement with the measurements of ref 10, although at much higher flow rate with a different, lager heat exchanger. Even after a long time of operating in this mode, the output line from the heat exchanger is at room temperature, and does not show any condensation of humidity.

The heat exchanger is mounted in a separate module on the cooling bus. It is convenient that it occupies the first position of modules since it requires an additional line originating below the liquid level in the detector. In the final version with two recirculation loops the Cooling Bus will be extended with a second heat exchanger module.

## 5. Emergency Cooling

In a worst case scenario cooling by the PTR would be interrupted either due to a general power failure, or malfunctioning of the cooling system. In such a case the xenon mass might be too large for quick recovery. The cooling bus, therefore, comprises a second cooling system based on $LN_2$. It will take over and keep the detector cooled for an extended period of time, either until a recovery or the return of the PTR cooling.

An additional module contains a stainless steel tube wound into a coil. $LN_2$ flowing through the coil condenses the xenon gas, and, the liquid will drop into a funnel in analogy to the PTR module. The PTR is regulated by a temperature controller, the $LN_2$ cooling, however, reacts to pressure increases. At a set pressure the $LN_2$ flow through the cooling coil will start and only stop after the pressure dropped below a second, lower set point. Since the $LN_2$ cooling will be used during emergencies its operation is entirely decoupled from the rest of the system. The pressure sensor, the controller, and the solenoid valve switching the $LN_2$ flow are powered by a small, dedicated UPS, and no other instrument will be permitted to share this power source. With the low consumption the battery of the UPS will last in excess of 24 hours.

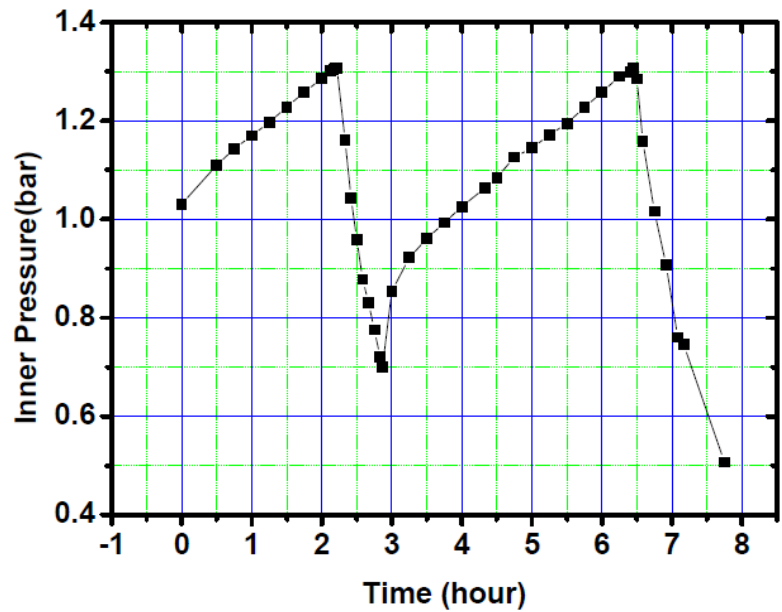

Fig. 5: Pressure in the inner vessel versus time showing the profile of slow warming up and rapid cool down with the $LN_2$ cooling system
.

The $LN_2$ cooling was tested with 100 kg of xenon in the detector vessel. The equilibrium state was established for a long time with the PTR at -94.5 °C and 1.03 BarG. The PTR compressor was then switched off, and there was no recirculation, but the pump on the insulation vacuum was still running. During an actual emergency this pump might have stopped, too. Therefore it is planned to back up the existing turbo molecular pump by a Dry Sorption pump. This kind of pump is cooled by $LN_2$ and does not require electricity.

The heat leaking into the system adiabatically evaporates some xenon, increasing the pressure, but keeping the equilibrium between gas pressure and liquid temperature. More than 2 hours past before reaching the set point at 1.3 BarG. After the solenoid valve opened it took about 5 minutes before the pressure started to decrease and another 40 minutes to reach the lower set point at 0.7 BarG. In the future the xenon mass will be 1000 kg instead of 100 kg. In a two hour period the amount of evaporated liquid will be similar, but the volume above the liquid will be smaller. Even when filled with 1000 kg the pressure will rise only to an estimated 1.7 BarG in a 2 hrs period. This is well below well below the 2.0 BarG the vessel is exposed to during every pre-cool cycle.

If the emergency has not been cleared during the initial 2 hrs, the $LN_2$ cooling will continue producing a pressure profile shown in fig. 5. The system is allowed to warm up for 200 minutes until the high pressure is reached and the cooled down in 40 minutes. This cycle will be repeated until the PTR will take over again.

## 6. Sensor Module

It is still an often heard misconception that the warm gas during filling must enter the system close to the PTR in order to be cooled. Practically, however, the gas can enter at any place. It will mix with the cold gas above the liquid, and the PTR will work on this mixture. In our detector the gas enters the system via the sensor module.

Other than being the filling port the sensor module houses all auxiliary sensors. In principle it is nothing else than an extension of the gas tube of the bus, bringing it outside of the vacuum insulation for easy installation of all the required gauges. The sensor module is shorter than regular modules, and has no gas flowing through it during normal operation. The length of stainless steel to any cold part in the cooling bus is very long, i.e. it is not a significant head load to the system.

In particular, the sensor module connects to 2 pairs of vacuum gauges, a Pirani and an Ion gauge, one for the inner vessel and one for the insulation vacuum. Additionally there is a pressure gauge to measure the system pressure during operation, a rupture disc set to 2.5 BarG as ultimate safety device, and the above mentioned fill line.

## 7. Filling and Emptying

For the present tests 100 kg of xenon were originally stored in gaseous form. Filling the detector is slow, limited by the available cooling power of the PTR. The rate depends very much on the heat capacity of the section to be cooled. Normally around 8 SLPM, the rate was reduced to only 2 SLPM while cooling down the heavy steel flange of the inner vessel. Once the liquid level was above the flange, the rate went back to 8 SLPM. The inner structure will reduce the rate only slightly due to the limited heat capacity of the components.

The average rate of 8 SLPM translates to about 2.5 kg/hr. This means the filling of 100 kg takes about 40 hrs. Future systems will have a much larger mass of xenon. To increase the flow we tested the concept of $LN_2$ assisted cooling. Without changing the settings of the PTR we flushed the emergency cooling coil with $LN_2$. As soon as the coil was cold we could increase the flow rate to 40 SLPM. The cooling of the PTR was reduced, and its heater was activated. This means that small variations in the LN2 flow would be compensated by the PTR. Several times we stopped the nitrogen flow, and reduced the fill rate. The switching between modes was very easy and without any incidents. By using the $LN_2$ cooling all the time the total amount of 100 kg could have been liquefied in about 8 hrs. This result can be even improved by a cooling coil with better heat transfer to the xenon gas.

To empty the vessel at the end of the run, the liquid was evaporated and then frozen in one of two large pressure cylinders (about 250 kg) cooled by $LN_2$. Well insulating this line the low vapor pressure above the frozen xenon in the storage tank would directly suck the liquid, without evaporating it. The positive pressure of around 1 BarG above the liquid would assist by pushing the liquid. The time required for emptying the detector would be drastically reduced, mainly limited by the flow characteristics of the connecting line. A control of the temperature in the storage tanks is not necessary, since finally it is irrelevant if the recuperated xenon is in liquid or solid form. The heat of xenon fusion is much smaller than the combined heat of fusion and liquefaction. The described liquid transport, therefore, reduces the $LN_2$ boil off by about 85% . In an underground lab this reduction is quite important, not only because of the logistic to provide the $LN_2$, but also because of the inherent dangers in releasing a large amount of nitrogen gas in a closed space. We want to point out that this mode of recovery can also proceed during a complete power failure and can be controlled remotely.

Before starting to fill the detector vessel, the system was pre-cooled. To this effect 2 BarG of Xenon were filled, and the cryo-cooler was started. This procedure is safe since even in the case of cooling failure the maximum pressure in the vessel cannot exceed 2 BarG. During pre-cooling we observed the effect of liquid drops being formed and falling on still hot surfaces. The pressure will rise at these events. These pressure spikes dreaded in small systems were completely harmless, and the maximum pressure rise was 30 mBar. The large volume of the vessel reduces the excess pressure from falling Xe drops. Note, however, that the forming of liquid already occurred during pre-cooling, about 2 hrs after cooling start. That means that the pre-cooling cycle also cools the connecting lines and finally the bottom of the vessel.

## 8. Conclusion

We tested the cryogenic system for the Panda-X liquid xenon detector in a laboratory environment. The geometry, e.g. vessel size, arrangement of modules, etc., corresponded to our plans for the final underground set up. The system was realized in a modular architecture which we called the 'Cooling Bus'. In this architecture modules can be added, replaced or removed as required without a complete re-design of the system.

A single 200 W PTR is sufficient even for 1 ton fiducial mass. With the services of a heat exchanger the cooling power is sufficient for a recirculation flow rate in excess of 100 SLPM, not limited by the PTR. With a smaller size pump we achieved 35 SLPM with an efficiency of the heat exchanger of 95.2%.

During tests with the emergency $LN_2$ cooling we measured the time for a pressure rise from 1 to 1.3 BarG of 2 hrs. This is due to the good thermal insulation of the system. We conclude that a future detector even with 1000 kg will not exceed a maximum pressure of 1.7 BarG within 2 hours. In case of any failure, sufficient time will be available to respond to the emergency.

We filled the detector with gaseous xenon at a rate of 8 SLPM. With $LN_2$ assisted cooling, a rate of 40 SLPM was reached. Even faster filling would be possible with a higher capacity $LN_2$ cooling coil.

### Acknowledgement

The Panda-X experiment is funded by a 973 project, No. 2010CB833005, of China's Ministry of Science and Technology, a special director's fund from the Natural Science Foundation of China, and a 985-III fund from Shanghai Jiao Tong University.
We would like to thank the other members of the Panda-X collaboration which contributed at various steps during the preparation of the test. We also thank Tom Haruyama from KEK, Tsukuba, Japan for many discussions during the design of the cryogenic system.